\documentclass[12pt,a4paper]{article}
\usepackage{keyval,graphicx}     

\begin{document}

\title{Energy Dependent Growth of Nucleon and Inclusive Charged Hadron Distributions\thanks
{Supported by Natural Science Foundation of Hebei Province
(A2012210043)}} \small{\date{}}
\author{\small {WANG Hong-Min$^{1}$\thanks{E-mail: whmw@sina.com.cn},
  , HOU Zhao-Yu$^{2}$, SUN Xian-Jing$^{3}$}\\
$^{1}$\small{Physics Department, Academy of Armored Forces
Engineering of PLA,
Beijing 100072, China}\\
$^{2}$\small{Physics Graduate School, Shijiazhuang Railway
Institute, 050043, China}\\
$^{3}$\small{Institute of High Energy Physics, Chinese Academy of
Sciences, Beijing 100049, China}}
 \maketitle
{\textbf{Abstract:}  In the Color Glass Condensate formalism,
charged hadron $p_{\mathrm{T}}$ distributions in p+p collisions are
studied by considering an energy-dependent broadening of nucleon's
density distribution. Then, in the Glasma flux tube picture, the
$n$-particle multiplicity distributions at different pseudo-rapidity
ranges are investigated. Both of the theoretical results show good
agreement with the recent experimental data from ALICE and CMS at
$\sqrt{s}=0.9,~2.36,~7$ TeV. The predictive results for
$p_{\mathrm{T}}$ and multiplicity distributions in p+p and p+Pb
collisions
at the Large Hadron Collider are also given in this paper.}\\



{\textbf{1. Introduction}}
\\

The measurement of charged hadron transverse momentum
($p_{\mathrm{T}}$) spectrum
 in heavy-ion collisions is a very significant topic on experimental physics.
Recently, the ALICE and CMS experiments at the Large Hadron Collider
(LHC) have given the data for charged hadron $p_{\mathrm{T}}$
distributions in p+p collisions at $\sqrt{s}=0.9,~2.36,~7$ TeV
[1-2]. The experimental results show that the scaling features of
charged hadrons previously observed at lower energies do not hold
good at high energies. These indicate that we still can not give
some basic understanding of the mechanism involved in the charged
particle production. Thus, a systematic theoretical study on this
topic is still very necessary.

An effective theory to study charged hadron $p_{\mathrm{T}}$
spectrum is the Color Glass Condensate (CGC) approach [3]. The CGC
is a state predicted by quantum chromodynamics (QCD) at high
energies where gluons in a hadron wave function expect to statute.
The cornerstone of the CGC approach is the existence of a hadron
saturation scale $Q_{s}$ at which gluon recombination effects start
to balance gluon radiation. In order to give an accurate theoretical
analysis, the nucleon's density distribution in position space must
be considered. Thus, the saturation scale, $Q_{s}$, should be
considered as a function of the nucleon's thickness function,
$T_{\mathrm{p}}(b)$. Furthermore, due to gluon saturation, the width
of the gluon distribution inside a nucleon should grow with
collision energy, $\sqrt{s}$~[4]. This will lead to a broadening of
the nucleon's density distribution in position space as $\sqrt{s}$
increases. Therefore, an energy-dependent broadening of the
nucleon's thickness function in the $Q_{s}$ should also be
considered.

Based on the CGC theory, many phenomenological saturation models,
such as the Golec-Biernat and W$\mathrm{\ddot{u}}$sthoff (GBW) model
[5], the Kharzeev, Levin and Nardi (KLN) model [6], the CGC model
[7] and the Kovchegov, Lu and Rezaeian (KLR) model [8], are
established. It should be noted that the KLR model is based on the
Anti-de Sitter space/conformal field theory (AdS/CFT). In this
paper, the KLN and KLR-AdS/CFT model are used because the simple
analytic unintegrated gluon distribution functions can be obtained
by a Fourier transform from the dipole amplitude of these two
models.

Integrating charged hadron transverse momentum  distributions over
$\mathbf{p}_{\mathrm{T}}$, we can obtain the impact parameter
(\textbf{b}) dependent mean multiplicity, $\bar{n}(\mathbf{b})$. The
mean multiplicity and the parameter controlling the size of the
fluctuations, $k$, are two parameters that characterize negative
binominal distribution (NBD) [9-10]. In the CGC picture, particles
produced locally in the transverse plane are considered as
correlating by approximately boost invariant flux tubes [11]. Since
there are $Q_{S}^{2}S_{\bot}$ such flux tubes emitting gluons, the
parameter $k$ should be proportional to $Q_{S}^{2}S_{\bot}$, where
$S_{\bot}$ is the transverse overlap area of two collision hadrons.
This indicates that the parameter $k$ in the NBD is also a function
of the thickness function, $T_{\mathrm{p}}(b)$, and influenced by
the energy-dependent broadening. By convolving the NBD and the
probability for an inelastic collision over the impact parameter,
the $n$-particle multiplicity distributions can be obtained.
\\

{\textbf{2. Energy Dependent Growth of Nucleon and Charged Hadron
$p_{\mathrm{T}}$ Distributions}}
\\

 In hadron-hadron collisions, the transverse momentum distributions for charged hadrons at leading order
can be expressed as [12]
$$\frac{dN(\mathrm{\textbf{b}})}{d^{2}\mathbf{p}_{T}}=\frac{1}{C_{F}}\frac{1}{p^{2}_{T}}\int
dy\int d^{2}\mathbf{r}_{T}
\int_{0}^{p_{T}}d^{2}\mathbf{k}_{T}\alpha_{s}(Q^{2})\phi_{1}(x_{1},
\frac{(\mathbf{k}_{T}+\mathbf{p}_{T})^{2}}{4},\mathbf{b})$$
\begin{equation}
\times\phi_{2}(x_{2},\frac{(\mathbf{k}_{T}-\mathbf{p}_{T})^{2}}{4},
\mathbf{b}-\mathbf{r}_{\bot}),
\end{equation}
where $N_{c}=3$, $x_{1,2}=(p_{_{T}}/\sqrt{s})\mathrm{exp}(\pm y),$
and $C_{F}=({N_{c}^{2}-1})/{(2\pi^{3}N_{c})}$. The running coupling
constant
$\alpha_{s}(Q^{2})=\mathrm{min}\{\frac{4\pi}{9\mathrm{ln}[{Q^{2}}/{\Lambda_{\mathrm{QCD}}^{2}}]},0.5\}$
with $\Lambda_{\mathrm{QCD}}=0.2$ GeV and
$Q^{2}=\mathrm{max}\{(\mathbf{k}_{T}+\mathbf{p}_{T})^{2})/4,(\mathbf{k}_{T}-\mathbf{p}_{T})^{2})/4\}$.
$\mathbf{b}$ and $\mathbf{r_{\bot}}$ are the impact factor and the
transverse position of the gluon, respectively.

 In Eq.(1), $\phi$ is the unintegrated gluon distribution (UGD).
 In the KLN model, $\phi$ can be taken as [6]
 \begin{equation}
 \phi(x,k^{2},\mathbf{b})=\frac{\xi C_{F}Q_{s}^{2}}{\alpha_{s}(Q_{s}^{2})}\left
 \{\begin{array}{cc}
 \frac{1}{Q_{s}^{2}+\Lambda^{2}}, &  k\leq Q_{s}\\
 \frac{1}{k^{2}+\Lambda^{2}}, &  k>Q_{s}
\end{array},
\right.
 \end{equation}
where $\xi$ is a normalization factor, and the saturation scale
[4-5]
\begin{equation}
Q_{s}^{2}(x)=Q^{2}_{0}(\frac{0.01}{x})^{\lambda},
\end{equation}
where $Q^{2}_{0}=2$ GeV$^{2}$, $\lambda=0.288$. In the KLR-AdS/CFT
model, $\phi$ can be obtained by a Fourier transform from the
dipole-nucleus amplitude given in Ref.[8]
\begin{equation}
\phi^{\mathrm{AdS}}(x,k^{2},\mathbf{b})=\int d^{2}\mathbf{r}
e^{i\mathbf{k}\cdot
\mathbf{r}}N(x,r)=\frac{32\pi}{(Q_{s}^{\mathrm{AdS}})^{2}}\frac{1}{[1+16k^{2}/(Q_{s}^{\mathrm{AdS}})^{2}]^{3/2}},
\end{equation}
where the corresponding saturation scale
\begin{equation}
Q_{s}^{\mathrm{AdS}}(x)=\frac{2A_{0}x}{M_{0}^{2}(1-x)\pi}(\frac{1}{\rho^{3}_{m}}+\frac{2}{\rho_{m}}-2M_{0}\sqrt{\frac{1-x}{x}}).
\end{equation}
The parameter $\rho_{m},M_{0}$ and $A_{0}$ in Eq.(5), as given in
Ref.[8], can be obtained by a fit to HERA data.

In order to give an accurate theoretical analysis, the UGD should be
considered as a function of the transverse position distribution of
the nucleon. Thus, the gluon saturation momentum is always written
as [4]
\begin{equation}
Q_{s,\mathrm{p(A)}}^{2}(x,\mathbf{b})=Q_{s}^{2}(x)(\frac{T_{\mathrm{p(A)}}(\mathbf{b})}{T_{\mathrm{p(A)},0}}),
\end{equation}
where the nuclear thickness
function
$$T_{\mathrm{p(A)}}(\mathbf{b})=\int dz\rho_{\mathrm{p(A)}}(\mathbf{b},z),$$
and $T_{\mathrm{p(A)},0}=1.53~\mathrm{fm}^{-2}$ [4]. For the nuclear
density distribution, we use the Woods-Saxon form [13]
\begin{equation}
\rho_{\mathrm{A}}(\mathbf{b},z)=\frac{\rho_{0}}{1+\mathrm{exp}[(r-R)/a]},
\end{equation}
where $r=\sqrt{\mathbf{b}^{2}+z^{2}}$, and the measured values for
Pb are $\rho_{0}=0.1612,R=6.62$ fm$,a=0.545$ fm. For the proton's
density distribution, the Gaussian form is used
\begin{equation}
\rho_{\mathrm{p}}(\mathbf{b},z)=\frac{e^{-r^{2}/(2B)}}{(2\pi
B)^{3/2}}.
\end{equation}
 Because of gluon saturation, the inelastic nucleon-nucleon cross
section $\sigma_{\mathrm{in}}$ should grow as $\sqrt{s}$ increases.
This will result in a broadening of the nucleon's density
distribution in position space. Therefore, the Gaussian width $B$
should be written as a function of $\sqrt{s}$ [4]
\begin{equation}
B(\sqrt{s})=\frac{\sigma_{\mathrm{in}}(\sqrt{s})}{14.30}\texttt{fm}^{2},
\end{equation}
with $\sigma_{\mathrm{in}}(\sqrt{s})=52,~60,~70.45,~72,~76.3$ mb at
$\sqrt{s}=0.9,~2.36,~7,~8.8,~14$ TeV, respectively [9,14-15].

In this paper, we also assume the gluon saturation scale should
 have a small dependence on $\sqrt{s}$ through the 3-dimensional
rms radius of the proton [16]
\begin{equation}
 Q^{2}_{s,\mathrm{p(A)}}(\sqrt{s})= Q^{2}_{s,\mathrm{p(A)}}(\sqrt{s_{0}})(\frac{\pi r^{2}_{\mathrm{rms},0}}{\pi
 r^{2}_{\mathrm{rms}}})^{1/\delta},
\end{equation}
where $\delta=0.8$ and the 3-dimensional rms radius
$r_{\mathrm{rms}}=\sqrt{<r^{2}>}=\sqrt{3B}$.
\\

{\textbf{3. Negative Binomial Distribution in the Glasma Flux Tube
Picture}}
\\

Negative binomial multiplicity distribution is an interesting and
important feature observed in multiplicity distribution of charged
hadron production in high energy collisions. In the Glasma flux tube
framework [11], the negative binomial distribution can be derived as
\begin{equation}
P_{n}^{\mathrm{NBD}}(\bar{n},k)=\frac{\Gamma(k+n)}{\Gamma(k)\Gamma(n+1)}(\frac{\bar{n}}{k})^{n}(1+\frac{\bar{n}}{k})^{-n-k},
\end{equation}
where the parameter $k$ and mean multiplicity $\bar{n}$  are all
considered as functions of the impact parameter.
 The parameter $k$ in the saturation approach is defined to be
\begin{equation}
k(\mathbf{b})=\zeta\frac{(N_{c}^{2}-1)Q_{s,\mathrm{p}}^{2}S_{\bot}(\mathbf{b})}{2\pi},
\end{equation}
where $\zeta$ is a dimensionless parameter and $S_{\bot}$ is the
overlap area of the two hadrons [11]. For a given impact parameter
\textbf{b},
\begin{equation}
Q_{s,\mathrm{p}}^{2}S_{\bot}(\mathbf{b})=\int
\mathrm{d}^{2}\mathbf{s}_{\bot}Q_{s,\mathrm{p}}^{2}(\mathbf{s}_{\bot},\mathbf{b}),
\end{equation}
where $Q_{s,\mathrm{p}}$ in the overlap area of the two hadrons is
chosen to be $Q_{s,\mathrm{p}}(\mathbf{s}_{\bot},\mathbf{b})=
\mathrm{min}\{Q_{s,\mathrm{p}}(\mathbf{s}_{\bot}),Q_{s,\mathrm{p}}(\mathbf{s}_{\bot}-\mathbf{b})\}$.
The mean multiplicity $\bar{n}$ can be obtained by integrating
Eq.(1) over $\mathbf{p}_{T}$
\begin{equation}
\bar{n}(\mathbf{b})=C_{\mathrm{m}}\int
d^{2}\mathbf{p}_{T}\frac{dN(\textbf{b})}{d^{2} \mathbf{p}_{T}},
\end{equation}
where the pre-factor $C_{\mathrm{m}}$ is proportional to
$\sigma_{\mathrm{in}}$.

In order to compute the probability distribution as a function of
multiplicity, one should convolve the NBD at a given impact
parameter (Eq.(11)) with the probability for an inelastic collision
($\frac{dP_{\mathrm{inel}}}{d^{2}\mathbf{b}}$) at the same impact
parameter. Thus, the probability distribution for producing $n$
particle can be given by
\begin{equation}
P(n)=\int d^{2}\mathbf{b} \frac{dP_{\mathrm{inel}}}{d^{2}\mathbf{b}}
P_{n}^{\mathrm{NBD}}(\bar{n}(\mathbf{b}),k(\mathbf{b})).
\end{equation}
where the probability distribution can be given in impact parameter
eikonal models [17]
\begin{equation}
\frac{dP_{\mathrm{inel}}}{d^{2}\mathbf{b}}=\frac{1-\mathrm{exp}(-\sigma_{\mathrm{gg}}T_{\mathrm{pp}})}
{\int
d^{2}\mathbf{b}[1-\mathrm{exp}(-\sigma_{\mathrm{gg}}T_{\mathrm{pp}})]},
\end{equation}
with the energy dependent quantity $\sigma_{\mathrm{gg}}=4\pi\lambda
B$ [4]. The overlap function for two protons at a given impact
parameter can be expressed as
\begin{equation}
T_{\mathrm{pp}}(\mathbf{b})=\int
d^{2}\mathbf{s}_{\bot}T_{\mathrm{p}}(\mathbf{s}_{\bot})T_{\mathrm{p}}(\mathbf{s}_{\bot}-\mathbf{b}).
\end{equation}
\\

\textbf{4. Results and Discussion}
\\

Charged hadron transverse momentum distributions in p+p collisions
at different collision energies are shown in Fig.1.  Fig.1 (a) and
(b) are the results with the KLN and the KLR-AdS/CFT model,
respectively. The $p_{\mathrm{T}}$ distributions are averaged over
the $\eta$ range from -2.4 to 2.4. The rapidity $(y)$ in Eq.(1) can
be changed into the pseudo-rapidity ($\eta$) using the
transformation
\begin{equation}
y(\eta)=\frac{1}{2}\mathrm{ln}\frac{\sqrt{\mathrm{cosh}^{2}\eta+m_{0}^{2}/p_{T}^{2}}+\mathrm{sinh}\eta}
{\sqrt{\mathrm{cosh}^{2}\eta+m_{0}^{2}/p_{T}^{2}}-\mathrm{sinh}\eta},
\end{equation}
where $m_{0}(\sim \mathrm{\Lambda}_{{\mathrm{QCD}}})$ is the rest
mass of particle. The experimental data come from CMS and ALICE
[1-2]. The parameter $\xi(=0.51)$ in Eq.(2) can be obtained by a
$\chi^{2}-$ analysis with the experimental data [18-19]. In Fig.1
(a), it is shown that the theoretical results with the KLN model are
in good agreement with the experimental data. For the KLR-AdS/CFT
model, as shown in Fig.1 (b), the theoretical results fit to the
experimental only at $\sqrt{s}=7$ TeV while $p_{\mathrm{T}}$ is
small. The reason is that the KLR-AdS/CFT model is an effective
model at small Bjorken-$x$ $(x<10^{-4})$.  At large $p_{\mathrm{T}}$
or small $\sqrt{s}$,
$x_{1,2}(=p_{\mathrm{T}}/\sqrt{s}\cdot\mathrm{exp}(\pm y))$ is out
of the valid range of this model. In Fig.2, the predictive results
with the KLN model for charged hadron $p_{\mathrm{T}}$ distributions
in p+Pb (solid curves) and p+p (dashed curves) collisions are also
given. Fig.2 (a) and (b) are the results at $\sqrt{s}=$5.02 TeV  and
8.8 TeV, respectively. It is shown that the results for p+Pb
collisions are closer to the values of p+p collisions at small
$p_{\mathrm{T}}$ than those at large $p_{\mathrm{T}}$.

Fig.3 (a) shows the the probability distribution for an inelastic
collision at $b$. The curves are the results at $\sqrt{s}=$ 0.9 TeV
(solid curve), 2.36 TeV (dashed curve), 7 TeV (dash-dotted curve),
and 14 TeV (dotted curve). The results show that the probability
distributions $2\pi b dP_{\mathrm{inel}}/d^{2}\mathbf{b}_{\bot}$ at
certain $\sqrt{s}$ have a sharply peaked distribution and the impact
parameter $b$ of the peak grows with increasing collision energy
$\sqrt{s}$. Fig.3 (b) shows the width parameter $k$ versus impact
parameter $b$. The figure captions are the same as those in Fig.3
(a). It is shown that the width parameter $k$ becomes larger as
$\sqrt{s}$ increases. This is because the number of flux tubes
$Q_{S}^{2}S_{\bot}$ in $k$ depends on the energy-dependent
broadening thickness function $T_{\mathrm{p}}$.

The results for charged hadron multiplicity distributions in the
range of $|\eta|<0.5$ and $|\eta|<1$ are shown in Fig.4. The figures
are the results at $\sqrt{s}=$ 0.9 TeV (a), 2.36 TeV (b), 7 TeV (c),
and 14 TeV (d). The solid curves are the results with the Glasma
flux tube approach, and the dashed curves are the results without
considering the nucleon's transverse position distribution. The data
come from ALICE [20]. The parameter $\zeta=0.05$ is extracted from a
fit to data at $\sqrt{s}=$ 2.36 TeV and used for all other energies.
  It is shown that the theoretical results with the
Glasma flux tube approach are in good agreement with the
experimental data at all collision energies. For the method without
considering the transverse distribution, the theoretical results fit
well to the data at $\sqrt{s}=$ 2.36 TeV but show deviations for
other collisions at highest multiplicities.

In summary, charged hadron $p_{\mathrm{T}}$ distributions in p+p
collisions at various collision energies are studied with the CGC
approach. By considering an energy-dependent broadening of the
nucleon's density distribution in position space, the theoretical
results of the KLN model are in good agreement with the experimental
data from CMS and ALICE. Then, the probability distribution for
producing $n$ particle at different pseudo-rapidity ranges are also
studied in the picture of Glasma flux tube, and the results fit well
to the data. The predictive results for $p_{\mathrm{T}}$ and
multiplicity distributions in p+p and p+Pb collisions will be
examined by the forthcoming experiment at the LHC.

\begin{newpage}

\end{newpage}

\begin{newpage}
\textbf{Figure Captions}\\

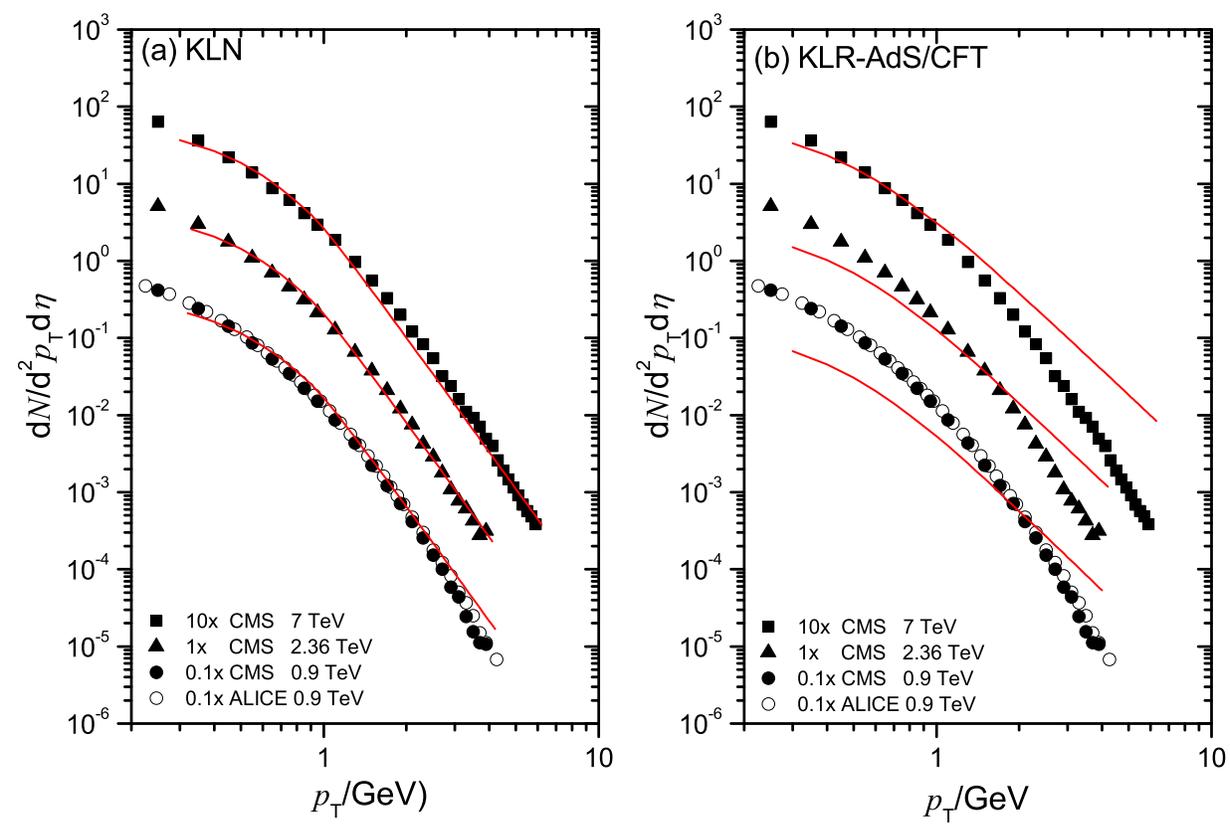
Figure1: Charged hadron $p_{\mathrm{T}}$ distributions for p+p
collisions averaged over the range of $|\eta|<2.4$ with the KLN
model (a) and the KLR-AdS/CFT model (b). The experimental data come
from CMS and ALICE [1-2].\\

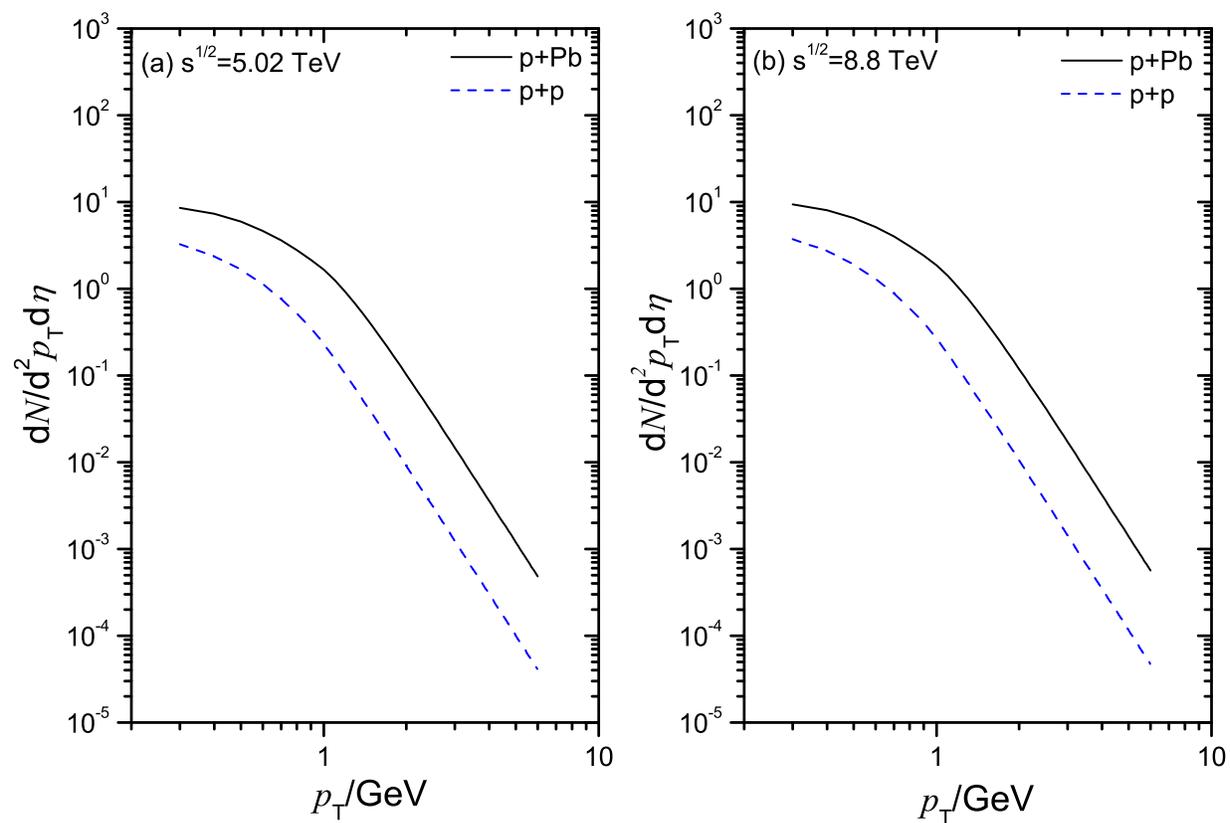
Figure2: The predictive results for charged hadron $p_{\mathrm{T}}$
distributions at $\sqrt{s}=$5.02 TeV (a) and 8.8 TeV  (b). The sold
and dashed curves are the results for p+Pb and p+p collisions,
respectively.\\

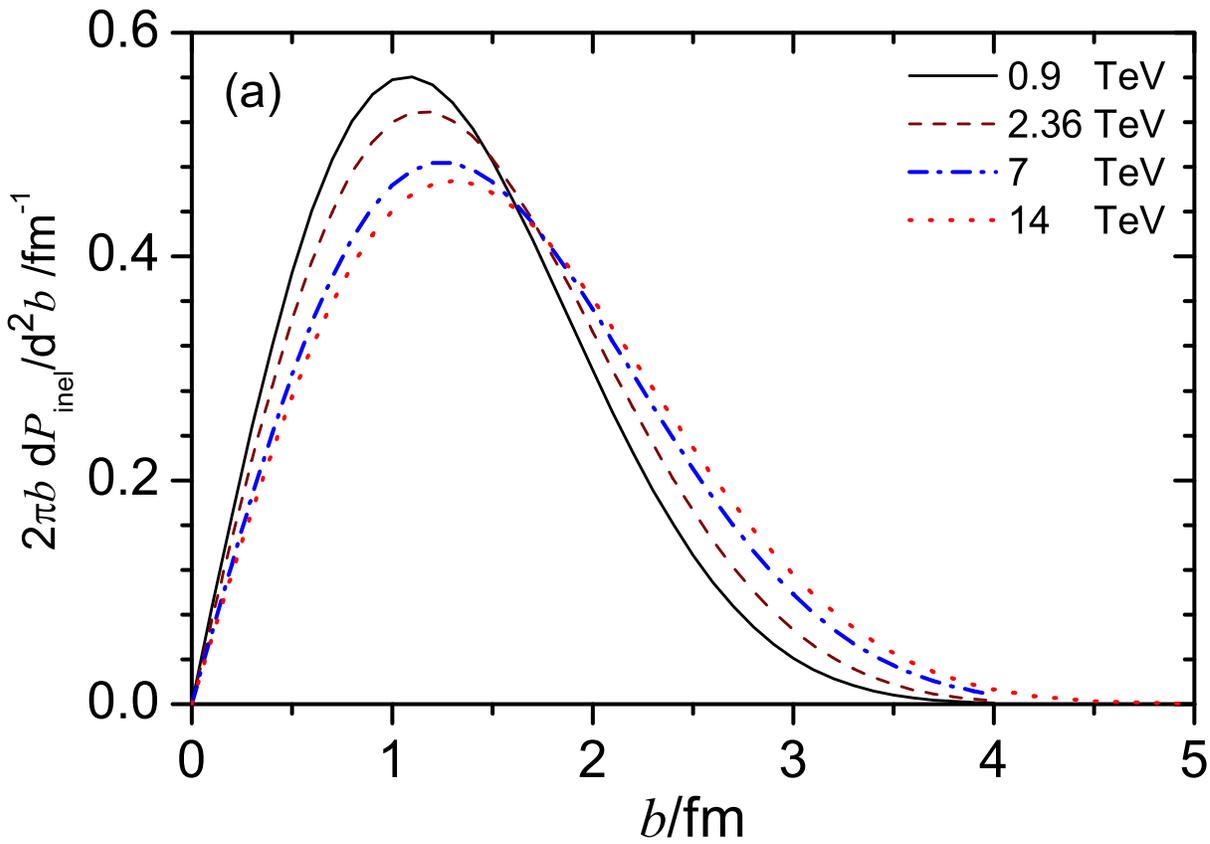
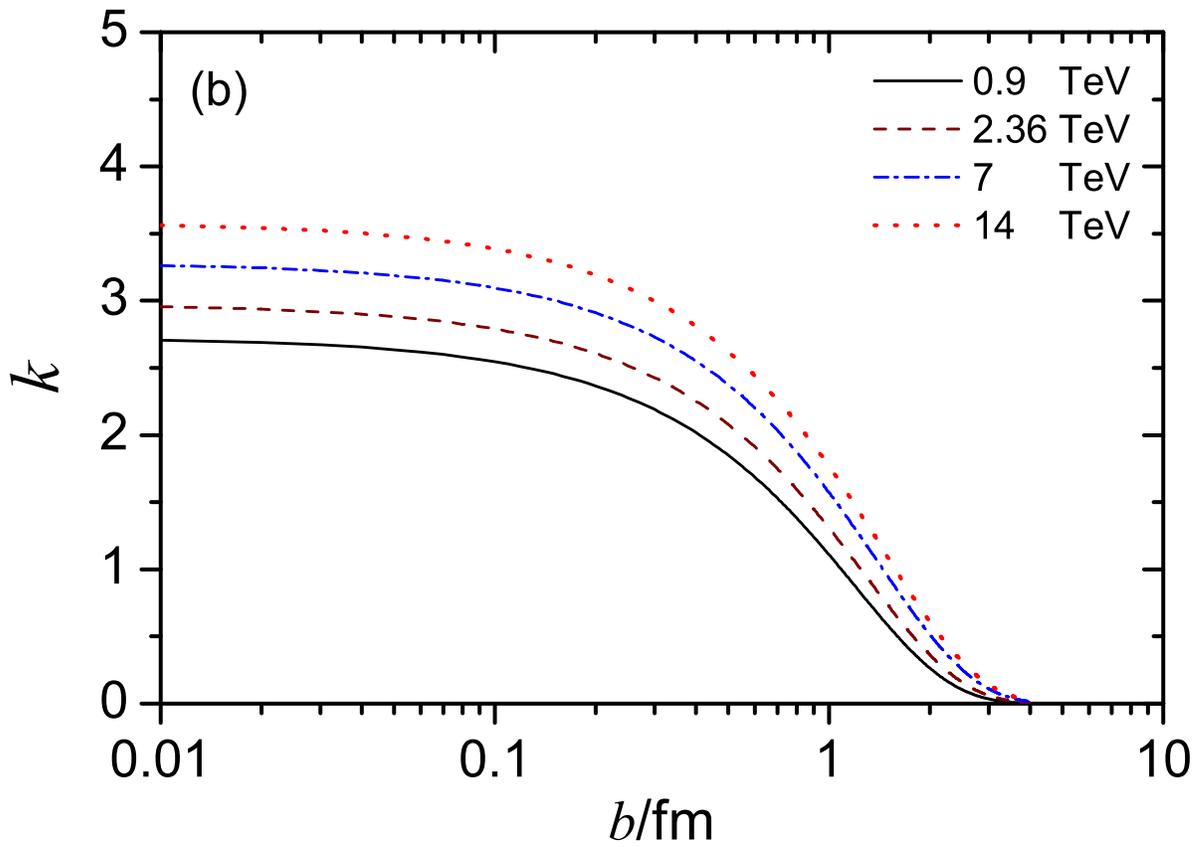
Figure3: The probability distribution for inelastic collision (a)
and the width parameter $k$ (b) as a function of impact parameter
$b$.\\

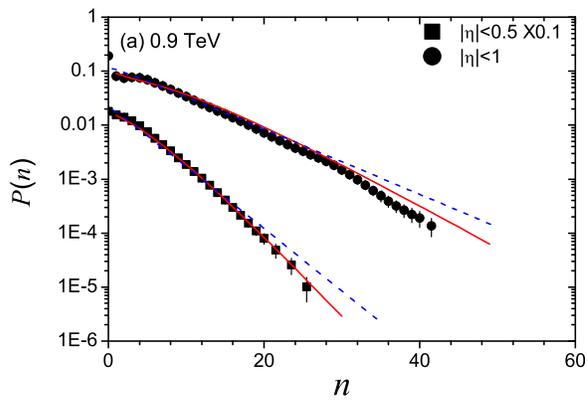
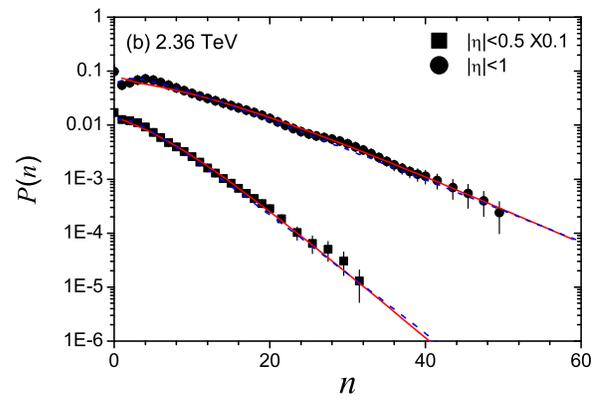
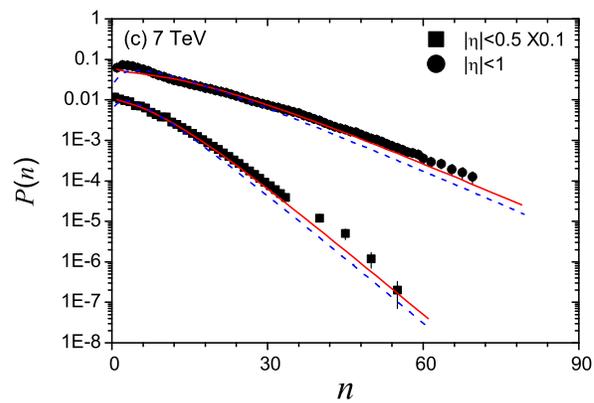
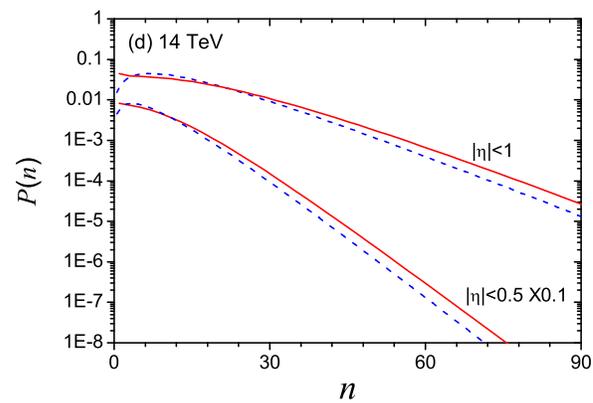
Figure4: Multiplicity distributions of charged hadrons in the range
of $|\eta|<0.5$ and $|\eta|<1$. The figures are the results at
$\sqrt{s}=0.9$ TeV (a) , 2.36 TeV (b), 7 TeV (c) and 14 TeV (d). The
solid curves are the results with the Glasma flux tube approach, and
the dashed curves are the results without considering nucleon's
position distribution. The data come from ALICE [20].

\end{newpage}
\end{document}